\documentclass[letterpaper]{article}
\usepackage{spconf,amsmath,amsfonts,amssymb}
\usepackage{graphicx}
\usepackage[hidelinks]{hyperref}
\usepackage{subfig}
\usepackage{xcolor}
\usepackage{braket}
\usepackage{float}
\usepackage{booktabs}
\usepackage{bm}
\usepackage{relsize}

\newcommand{\bigTau}{\mathlarger{\mathlarger{\tau}}}

\title{
Characterizing the Effects of Mixed Division Waveform Schemes in MIMO Radar
}
\name{Oliver P.R. Kirkpatrick$^{1}$, Santiago Ozafrain$^{1}$, Christopher Gilliam$^{2}$, Beth Jelfs$^{1}$}
\address{$^{1}$School of Engineering, University of Birmingham, UK\\
$^{2}$Department of Electrical Engineering and Electronics, University of Liverpool, UK\\
Email: opk323@student.bham.ac.uk, \{s.ozafrain, b.jelfs\}@bham.ac.uk, c.gilliam@liverpool.ac.uk}

\begin{document}
\ninept % things fit really nicely
\maketitle 
%=======================================================================
\begin{abstract}
Direction of departure estimation in Multiple-Input-Multiple-Output (MIMO) radar is based on transmitting a set of orthogonal (or at least separable) waveforms, one per transmit element. Assuming orthogonality is preserved, these waveforms can then be separated into different channels at the receiver before beamforming. In practice, however, MIMO waveforms are not perfectly orthogonal, and at different points in the processing chain, this can manifest as phase errors in the steering manifolds used in beamforming. Furthermore, due to resource constraints, either by the channel or the sensing scenario, a mixture of orthogonality techniques, \textit{mixed division strategies}, are often employed, resulting in further phase errors. Naturally, such errors will interfere in the beamforming process, spoiling the beam, introducing direction finding errors, or increasing sidelobe levels, and consequently reducing SNR. In this paper, we demonstrate how mixing division strategies significantly degrades orthogonality of otherwise orthogonal waveforms resulting in degraded beamforming performance. Specifically, we show that waveforms which exhibit favourable sidelobe levels may still induce large direction of departure errors and vice versa. Additionally, through multiple exhaustive searches of potential mixed division strategies with combinations of time-, Doppler- and code-based division techniques we show that large numbers of the possible waveforms exhibit poor beamforming qualities.
\end{abstract}

%=======================================================================

%=======================================================================


%=======================================================================

%=======================================================================
\section{Introduction}
\label{sec:Intro}
%=======================================================================

The design of multiple-input-multiple-output (MIMO) systems remains an active area of research for a wide variety of radar applications~\cite{bergin2018mimo,overdestdoppler,frazer2007spatially,6558024,abramovich2015aperiodicwaveforms}. The popularity of MIMO arises due to several factors: increased clutter rejection capability~\cite{abramovich2015aperiodicwaveforms}; spatial diversity~\cite{frazer2007spatially,molaei2025efficient}; and support for anti-jamming and interference rejection~\cite{molaei2025efficient,xue2024optimized}. The extent to which a radar system can utilize these capabilities depends heavily upon array geometry, antenna gain pattern~\cite{bergin2018mimo,kirkpatrick2026mimoarraycalibrationnonstationary}, and waveform design~\cite{overdestdoppler}. In particular, optimal MIMO waveform design enables transmit beamforming to be performed in any and all directions simultaneously on receive. However, successful realization of on receive beamforming requires the signals transmitted by each channel to be separable. Typically, this is achieved by ensuring the waveforms are mutually orthogonal to one another~\cite{overdestdoppler}. Therefore, as the MIMO radar community attempts to extract ever-increasing levels of performance out of their systems, the number of mutually orthogonal waveforms required grows, inevitably exposing the shortcomings of any single waveform design~\cite{overdestdoppler,sun2014analysis}.

In general, MIMO waveforms are based on Division Multiple Access (DMA) techniques where mutual orthogonality between the transmit channels is achieved using divisions in time (TDMA), frequency (FDMA), Doppler velocity (DDMA), or via codes (CDMA)~\cite{sun2014analysis}, with each division scheme having its own shortcomings. Time division schemes, popular for their ease of hardware implementation~\cite{xu2021transmit}, can reduce unambiguous range. Doppler division schemes can preserve unambiguous range, at the expense of reducing maximum unambiguous Doppler velocity. Frequency division can alleviate both limitations, but introduces complexity and range coupling in direction of departure (DoD) estimation, and is restricted by the channel coherence bandwidth. Applying mutually orthogonal codes in slow-time is sensitive to target velocity; and fast-time codes exhibit increasingly poor sidelobe levels for each additional channel. While, much effort has been put into alleviating the limitations of each scheme~\cite{xu2021transmit,hong2024multiple,xie2025intelligent}, Mixed DMA (MDMA), also called joint coding~\cite{liu2026stepped}, aims to maximize the number of MIMO channels which can be supported, without impacting the radar performance, by mixing multiple schemes together~\cite{liu2026stepped,wang2023doppler,liu2025range,chen2026dual}. Naturally, as more division schemes are combined, it is to be expected this will bring its own undesirable effects.

In this paper, we demonstrate that mixed division schemes introduce significant errors to the MIMO beamforming process that are not otherwise present when the component division schemes are used in isolation. We focus on MDMA schemes involving permutations of TDMA, DDMA, and CDMA~\cite{wang2023doppler,liu2026stepped,chen2026dual,pypalli2018orthogonal,yasser2025precise}. We show that errors in beamforming can be prevalent across a wide range of possible codes in the parameter space of a selected CDMA scheme (Zadoff-Chu sequences), making many of the possible resultant MDMA combinations undesirable. We propose that these waveforms are better evaluated using a full radar simulation of pseudorandom targets, than with simple comparisons of waveform cross correlation, due to the range and direction of departure dependency of the errors.

This paper is structured as follows: \autoref{sec:signal-model} introduces the signal model and processing pipeline for MIMO radar and beamforming; \autoref{sec:MIMO_Waveforms} introduces selected division schemes; \autoref{sec:performance-analysis} details an exhaustive search of different MDMA schemes, to give insight into where orthogonality begins to break down in MDMA MIMO; finally, \autoref{sec:conclusion} concludes the paper.

%=======================================================================
\section{MIMO Beamforming}
\label{sec:signal-model}
%=======================================================================

To simulate a MIMO system of $M_{t}$ transmit and $M_{r}$ receive channels, we use a baseband simulation method, and employ slow-time Doppler processing to accomplish coherent integration. To remove potential confounding variables, we simulate only stationary targets\footnote{The influence of the target's velocity on its ambiguity function has previously been established~\cite{overdestdoppler}} and consider a noiseless environment, and unity gain and power. Each receive channel is modelled as the sum of the transmitted signals, $\left\{s_{m_{t}}\left(t\right)\right\}_{m_t=0}^{M_t-1}$, with each transmit channel incurring a time delay that reflects its unique path to the receive element. Thus, for the $p^{th}$ pulse, the $n^{th}$ sample of the $m_{r}^{\text{th}}$ received channel is:
\begin{equation}\label{eq:signal-model}
    x_{m_{r}}\left[p,n\right]=\sum_{m_t=0}^{M_{t}-1}s_{m_{t}}\left(\frac{n}{f_{s}}-\bigTau_{p,m_{t},m_{r}}\right)\operatorname{e}^{-2\pi jf_{c}\bigTau_{p,m_{t},m_{r}}}
\end{equation}
where $\bigTau_{p,m_{t},m_{r}}$ is the delay of the signal from the $m_{t}^{\text{th}}$ transmitter to the $m_{r}^{\text{th}}$ receive element on the $p^{th}$ pulse. The carrier frequency of the system is defined by $f_c$ and the sampling frequency by $f_s$.

Coherent integration of a set number of pulses or waveforms is accomplished via range compression and slow-time Doppler processing. For each transmit channel, range compression is applied via correlation with a local replica of the transmitted signal, $s_{m_{t}}$:
\begin{equation}
    y_{m_{r},m_{t}}\left[p,r\right]=\sum_{n=0}^{N-1}x_{m_{r}}\left[p,n\right]s_{m_{t}}\left[\left(n-r\right)\;\operatorname{mod}\;N\right]^{\ast},
\end{equation}
followed by Doppler processing across the slow-time dimension, yielding range-Doppler data:
\begin{equation}
    Y_{m_{r},m_{t}}\left[k,r\right]=\sum_{p=0}^{P-1}y_{m_{r},m_{t}}\left[p,r\right]\operatorname{e}^{-2\pi j\frac{k}{P}p}.
\end{equation}
In this formulation, $P$ denotes the number of transmitted pulses, $N$ the number of samples within the pulse repetition interval, $k$ denotes the Doppler frequency index and $r$ the delay index. Thus, after full channel separation, there are $M_{t}\times M_{r}$ total range-Doppler outputs. Importantly, depending on the chosen division strategy, only a subregion of the generated $Y_{m_{r},m_{t}}$ should be retained for each MIMO channel; MDMA schemes are employed to maximise both this subregion and the total number of transmit-receive channels.

\begin{figure}[t]
    \centering
    \includegraphics[width=0.6\linewidth]{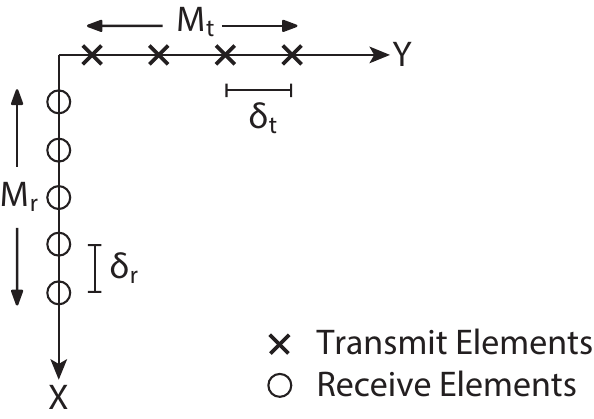}
    \caption{Geometry of orthogonal and co-located uniform linear transmit and receive arrays.}
    \label{fig:array-geometry}
\end{figure}

\begin{figure}
    \centering
    \includegraphics[width=0.99\linewidth]{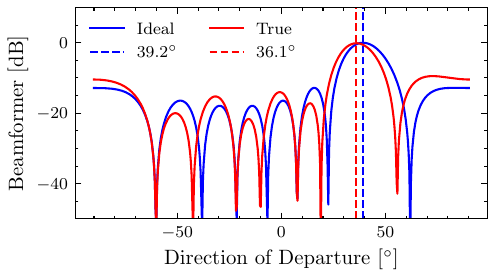}
    \caption{Comparison of an ideal direction of departure beampattern (blue), and beampattern formed when using imperfectly orthogonal waveforms (red).}
    \label{fig:example-beamformer-output}
\end{figure}

Finally, MIMO beamforming is performed by approximating the phase shifts across the transmit and receive arrays as simple functions of azimuth, $\vartheta$, and elevation, $\varphi$. Assuming the array geometry shown in~\autoref{fig:array-geometry}, for transmit and receive array inter-element spacings of $\delta_{t}$ and $\delta_{r}$, respectively, the MIMO steering vector is given by~\cite{kirkpatrick2026mimoarraycalibrationnonstationary}:
\begin{align}\label{eq:mimo-steering-vec}
    \mathbf{a}\left(\vartheta,\varphi\right)=&\begin{bmatrix}
        1\\
        \exp\left(-2\pi j\frac{m\delta_{r}}{\lambda}\cos\vartheta\cos\varphi\right)\\
        \vdots\\
        \exp\left(-2\pi j\frac{\left(M_{r}-1\right)\delta_{r}}{\lambda}\cos\vartheta\cos\varphi\right)
    \end{bmatrix} \nonumber\\
    &\otimes\begin{bmatrix}
        1\\
        \exp\left(-2\pi j\frac{m\delta_{t}}{\lambda}\sin\vartheta\cos\varphi\right)\\
        \vdots\\
        \exp\left(-2\pi j\frac{\left(M_{t}-1\right)\delta_{t}}{\lambda}\sin\vartheta\cos\varphi\right)
    \end{bmatrix},
\end{align}
where $\lambda$ is the wavelength corresponding to the carrier frequency and $\otimes$ is the Kronecker product. For a given range-Doppler cell, MIMO beamforming is achieved by computing the inner product of that cell across all $M_{t}M_{r}$ channels with~\eqref{eq:mimo-steering-vec}. Successful MIMO beamforming theoretically amplifies the response from a target by $M_tM_r$. However, the ability to achieve this gain is hampered by any deviation of the target's response from the ideal steering vector defined in \eqref{eq:mimo-steering-vec}. One source of these deviations is the phase errors induced by loss of orthogonality across the MIMO waveforms. Such phase errors can act to spoil the beam by reducing null depth or increasing the sidelobe level, or biasing estimated direction of departure~\cite{overdestdoppler} as illustrated in \autoref{fig:example-beamformer-output}. 

%=======================================================================
\section{Common MIMO Multiple Access Schemes}
\label{sec:MIMO_Waveforms}
%=======================================================================

In this section, we outline the MIMO division strategies that will be used in our MDMA study. We restrict ourselves to division strategies based on manipulating a prototype chirp waveform:
\begin{equation}\label{eq:chirp}
    s(t) = \exp\left(2\pi j\left(f_0 t + \frac{1}{2}\alpha t^2\right)\right),
\end{equation}
where $\alpha$ defines the chirp rate. For a more general description of MIMO waveforms, we refer the reader to~\cite{sun2014analysis}. \vspace{1.5mm}

\noindent\textbf{Time Division (TDMA)} applies an unique time delay, $m_{t}\bigTau_{\text{td}}$, to each of the transmitted signals such that $s_{m_t}(t) = s(t - m_{t}\bigTau_{\text{td}})$. The effect in the delay-Doppler space is that target responses are tessellated $M_{t}$ times along the delay dimension. If the time delay is greater or equal to the waveform duration, $T$, then the set of waveforms are strictly orthogonal~\cite{sun2014analysis,chen2026dual}. However, enforcing this condition reduces the unambiguous Doppler frequency the system can measure. An alternative approach is time staggered TDMA (TS-TDMA) where the time delay is $m_t T/M_t$. A drawback of this approach is that the individual transmit channels now overlap in time. These signals can still be separated using appropriate matched filtering however the tessellation in the delay dimension will overlap, which reduces the unambiguous range of the system. Thus, use of TS-TDMA or TDMA poses a simple trade-off unambiguous range vs unambiguous Doppler frequency, with increasing number of channels affecting one, or the other or both. \vspace{1.5mm}

\noindent\textbf{Doppler Division (DDMA)} is a slow-time coding method that applies a unique phase modulation, $\phi_{m_t,p}$, per waveform per transit element~\cite{xie2025intelligent} such that $s_{m_t}(t) = s(t)\operatorname{e}^{j\phi_{m_t,p}}$. The modulation is defined such that across $P$ transmitted pulses a pseudo Doppler frequency is inducted that is unique to that transmit element. The effect in the delay-Doppler space is that target responses are tessellated $M_{t}$ times along the Doppler dimension. These responses can be easily separated using appropriate Doppler processing however the maximum/minimum unambiguous Doppler frequency is reduced; this reduction is proportional to the number of transmit channels. \vspace{1.5mm}

\noindent\textbf{Code Division (CDMA)} is a fast-time coding method that applies a unique multi-symbol code, $\phi_{m_t}(t)$, per transmit element such that $s_{m_t}(t) = s(t)\operatorname{e}^{j\phi_{m_t}(t)}$. Notionally, the number of possible transmit channels CDMA can support is dependent on how many unique, orthogonal, codes it can generate. The number of codes is governed by the chip size --- measured in samples --- of the symbols used in the code; smaller chips allow more codes. There exists numerous ways of generating symbols, such as pseudorandom codes~\cite{rabideau2012mimo}, Gold Code sequences~\cite{zepernick2013pseudo}, and Almost Perfect Autocorrelation Sequences (APAS)~\cite{144729}. For this paper, however, we focus on Zadoff-Chu sequences~\cite{budisin2010decimation} due to their established use in MIMO radar and computationally tractability.

%=======================================================================
\section{Analysis of Waveform Orthogonality Loss}
\label{sec:performance-analysis}
%=======================================================================

In this section we analyse the orthogonality of different MDMA schemes. For all simulations, all targets are stationary, and signals are simulated in the absence of noise, we assume 8 transmit channels and 1 receive channel, with a chirp bandwidth of 1/1000 of the carrier frequency, and a sample rate of 5/1000 of the carrier ($f_{c}=10$ MHz). 128 total pulses were integrated, with each pulse consisting of 1024 samples.

For each given division scheme, we simulate 100 targets with pseudorandom (unambiguous) ranges and directions of departures, spanning $\pm45^{\circ}$ azimuth. To gauge the impact of the waveform orthogonality on the beamformer performance we calculate the RMSE over the 100 targets of the error between the target's true direction of departure and the estimated DoD, denoted $\text{DoD RMSE}$. Similarly, under the assumption of a single target, we calculate the the median sidelobe level (MSLL) by determining the power of the largest sidelobe for each MIMO channel $Y_{m_{r},m_{t}}$ and then computing the median across the $M_tM_r$ MIMO channels and finally averaging across the 100 targets to give $\overline{\text{MSLL}}$. 

First, to establish a baseline performance, we test TDMA and DDMA, individually\footnote{Note that throughout these simulations we use TS-TDMA}. Results for the direction of departure RMSE and the MSLL averaged across the 100 simulated targets are shown in \autoref{tab:baseline-beamformer-quality-loss}. The table also includes illustrative MDMA schemes using combinations of TDMA and DDMA. From \autoref{tab:baseline-beamformer-quality-loss}, it can be observed that while the sidelobe performance is approximately the same across waveforms, schemes incorporating TDMA exhibit greater DoD error, approximately proportional to the number of TDMA groups. These values suggest that TDMA is intrinsically prone to degradation as more channels are introduced. 

\begin{table}[tb]
    \centering
    \caption{Direction of departure errors and sidelobe levels for baseline TDMA and DDMA orthogonal waveforms along with MDMA schemes consisting of combinations of the two.}
    \label{tab:baseline-beamformer-quality-loss}
    \begin{tabular}{lcc}
        \toprule
        Division Strategy & DoD RMSE [$^{\circ}$] & $\overline{\text{MSLL}}$ [dB]\\
        \midrule
        $8\times$ TDMA & $0.0763$ & $-17.03$ \\
        $8\times$ DDMA & $0.0037$ & $-17.95$\\
        $2$ TDMA $\times$ $4$ DDMA & $0.0116$ & $-17.96$ \\
        $4$ TDMA $\times$ $2$ DDMA & $0.0345$ & $-17.70$ \\
        \bottomrule
    \end{tabular}
\end{table}

\begin{table*}[tb]
    \centering
    \caption{Direction of departure errors and sidelobe levels for different MDMA schemes which implement CDMA. The results show median values with interquartile ranges of the metrics after an exhaustive search of pairs of Zadoff-Chu codes. U/D denotes MDMA schemes that use up and down chirps for the Zadoff-Chu codes.}
    \begin{tabular}{@{\hspace{0.1em}}lc@{\hspace{0.6em}}c@{\hspace{0.6em}}c@{\hspace{1.4em}}c@{\hspace{0.6em}}c@{\hspace{0.6em}}c@{\hspace{1.4em}}c@{\hspace{0.6em}}c@{\hspace{0.6em}}c@{\hspace{0.1em}}}
        \toprule
        Division & \multicolumn{3}{c}{DoD RMSE $\left[^{\circ}\right]$} & \multicolumn{3}{c}{$\overline{\text{MSLL}}$ [dB]} &  \multicolumn{3}{c}{Usable}\\
        Strategy & \multicolumn{3}{c}{Median (IQR)} & \multicolumn{3}{c}{Median (IQR)} &  \multicolumn{3}{c}{Fraction [\%]}\\
        \cmidrule{2-10}
        \multicolumn{1}{r}{$N_{ZC}=$} & $8$ & $16$ & $32$ & $8$ & $16$ & $32$ & $8$ & $16$ & $32$ \\
        \midrule
        CDMA, TDMA & $0.14\ (2.74)$ & $0.13\ (1.83)$ & $0.27\ (1.21)$ & $-6.89\ (2.08)$ & $-6.90\ (1.95)$ & $-6.07\ (2.75)$ & $0.59$ & $0.54$ & $0.55$\\
        CDMA, DDMA & $0.03\ (2.82)$ & $0.03\ (1.91)$ & $0.03\ (1.36)$ & $-7.00\ (2.13)$ & $-7.16\ (2.13)$ & $-6.62\ (3.05)$ & $0.88$ & $1.85$ & $3.72$\\
        CDMA, TDMA-DDMA & $0.04\ (2.81)$ & $0.05\ (1.89)$ & $0.04\ (1.34)$ & $-6.98\ (2.13)$ & $-7.09\ (2.09)$ & $-6.56\ (2.99)$ & $0.88$ & $1.63$ & $3.25$\\
        \midrule
        U/D CDMA, TDMA        & $0.47\ (0.21)$ & $0.37\ (0.17)$ & $0.57\ (0.28)$ & $-7.97\ (0.48)$ & $-6.86\ (1.58)$ & $-6.11\ (2.81)$ & $0.00$ & $0.00$ & $0.56$\\
        U/D CDMA, DDMA        & $0.20\ (0.14)$ & $0.16\ (0.10)$ & $0.25\ (0.18)$ & $-10.1\ (0.60)$ & $-8.54\ (2.18)$ & $-7.33\ (3.71)$ & $54.2$ & $22.9$ & $15.6$\\
        U/D CDMA, TDMA-DDMA   & $0.26\ (0.10)$ & $0.21\ (0.10)$ & $0.33\ (0.17)$ & $-9.36\ (0.57)$ & $-7.78\ (1.94)$ & $-6.67\ (3.31)$ & $5.86$ & $9.30$ & $4.51$\\
        \bottomrule
    \end{tabular}
    \label{tab:beamformer-quality-loss}
\end{table*}

\begin{figure}[tb]
    \centering
    \includegraphics[width=0.99\linewidth]{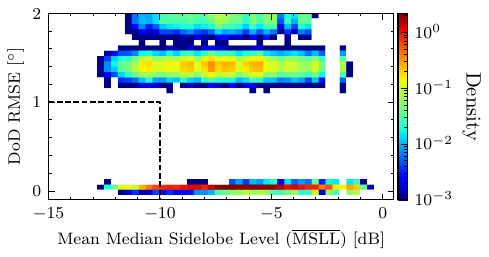}
    \caption{2D Histogram of DoD RMSE and $\overline{\text{MSLL}}$ for an exhaustive search of pairs of Zadoff-Chu codes for an MDMA strategy with $2$ TDMA, $2$ DDMA, and $2$ CDMA ($N_{ZC}=32$) groups. The dashed black line illustrates a threshold region with $\leq1^{\circ}$ of DoD RMSE and $\overline{\text{MSLL}}\leq10$ dB.}
    \label{fig:parameter-space-search}
\end{figure}

We now go on to show that this trend of TDMA eroding beamforming performance is only exacerbated by introduction of CDMA. MDMA schemes which include CDMA were implemented using pairs of Zadoff-Chu codes embedded on chirps, so as to halve the number of TDMA and/or DDMA groups required. We expect that any given pair of Zadoff-Chu codes will interact with other division schemes in unique ways, resulting in MDMA schemes that exhibit a wide variety of sidelobe levels and direction of departure errors, some of which may be unacceptable. Accordingly, we evaluate schemes using codes with varying sequence lengths ($N_{ZC}=8,16,32$), and for each sequence length, perform an exhaustive search of the full range of possible Zadoff-Chu pairs. As the full waveform consists of 1024 samples, each $N_{ZC}=8,16,32$ length sequence has a chip size of $128$, $64$, and $32$, respectively. 

An illustrative example of the outputs of such a search is shown in \autoref{fig:parameter-space-search} for an MDMA scheme comprising two Zadoff-Chu ($N_{ZC}=32$) CDMA groups, two TDMA groups, and two DDMA groups (i.e., $M_{t}=8$ in total). The figure shows the 2D histogram of the DoD RMSE and $\overline{\text{MSLL}}$ for all the possible combinations of Zadoff-Chu code pairs. Statistical values for all of the possible MDMA combinations are presented in \autoref{tab:beamformer-quality-loss}. The table details the median values and interquartile ranges for the DoD RMSE and $\overline{\text{MSLL}}$ metrics obtained by exhaustively searching all possible combinations of Zadoff-Chu code pairs.

There are two main observations to take from the histogram in \autoref{fig:parameter-space-search}. The first is that a low sidelobe level does not guarantee an accurate direction of departure estimate and vice versa. The second is that the distribution is multimodal. This multimodal distribution gives rise to the large interquartile range values presented in \autoref{tab:beamformer-quality-loss}. The table shows that for a given MDMA scheme the median values for the DoD RMSE and $\overline{\text{MSLL}}$ are relatively consistent when changing code sequence length $N_{ZC}$ however interquartile ranges are not. More generally, the results in the table display the same trend as those for the baseline schemes in \autoref{tab:baseline-beamformer-quality-loss}; introduction of TDMA groups degrades the performance of the radar system and the degradation is proportional to the number of TDMA groups used. The table also shows that increasing the sequence length decreases the interquartile range for the DoD RMSE for all of the schemes. 

To evaluate what this means for a practical system, we define ``usable'' waveforms based on a criteria of DoD RMSE $\leq1^{\circ}$ and $\overline{\text{MSLL}}\leq10$ dB. Note that this criteria is very application dependent. Using this criteria, we obtain the fraction of the total number of waveforms that are actually usable. From \autoref{tab:beamformer-quality-loss}, we can see that only a small fraction of the possible Zadoff-Chu code pairs satisfy our criteria. This result suggests that design of MDMA waveforms cannot be reduced to just ensuring the individual groups are orthogonal.  

\begin{figure}[tb]
    \centering
    \includegraphics[width=0.99\linewidth]{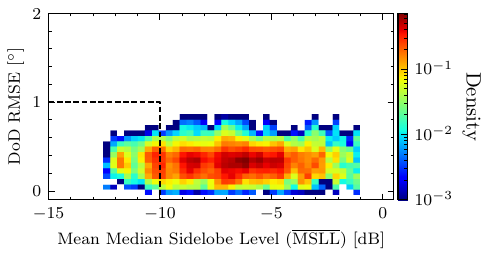}
    \caption{2D Histogram of DoD RMSE and $\overline{\text{MSLL}}$ for an exhaustive search of pairs of Zadoff-Chu codes for an MDMA strategy with $2$ TDMA, $2$ DDMA, and $2$ CDMA ($N_{ZC}=32$), where one code is embedded on an up chirp, and the other on a down chirp. The dashed black line illustrates a threshold region with $\leq1^{\circ}$ of DoD RMSE and $\overline{\text{MSLL}}\leq10$ dB.}
    \label{fig:parameter-space-search-ud}
\end{figure}

Finally, we broaden our investigation to consider what happens when embedding the Zadoff-Chu codes on chirps with differing chirp rates. Specifically, we consider up and down chirps created by changing the sign of the chirp rate; although not strictly orthogonal, such an approach has seen increased attention due to favourable properties in poor channels~\cite{pypalli2018orthogonal,yasser2025precise}. Similar to the previous results, the statistics for the exhaustive Zadoff-Chu search are show in \autoref{tab:beamformer-quality-loss} and an example 2D histogram of the DoD RMSE and $\overline{\text{MSLL}}$ for all the possible combinations of Zadoff-Chu code pairs is shown in \autoref{fig:parameter-space-search-ud}. Note that the MDMA scheme tested in \autoref{fig:parameter-space-search-ud} is equivalent to the scheme in \autoref{fig:parameter-space-search}. Both the figure and the table indicate a more concentrated distribution in performance metrics, which translates in smaller interquartile range values. More generally, the table shows that the median $\overline{\text{MSLL}}$ is generally reduced when using up and down chirps and the reverse is true for the median DoD RMSE. However, the increase in median DoD RMSE is relatively small, which is demonstrated by the increase in the percentage of usable waveforms based on our criteria.  

%=======================================================================
\section{Conclusion}
\label{sec:conclusion}
%=======================================================================

In this paper, we have investigated the performance of MDMA waveform schemes for MIMO radar. Specifically, we have focused on MDMA schemes that seek to jointly maximise the number of transmit channels and the unambiguous range-Doppler space of the MIMO radar. Through full radar simulations, we have demonstrated that, even in noiseless conditions, MDMA schemes constructed from groups of orthogonal waveforms in time, Doppler and/or code-space can exhibit large DoD errors and large sidelobe levels. Furthermore, we showed that low sidelobe levels does not translate to small DoD errors and vice versa. In particular, use of TDMA resulted in greater DoD error than other DMA schemes. Accordingly, a key outcome from our investigation is that the design of a MDMA scheme from combinations of CDMA, TDMA and DDMA needs to consider the interactions between all of the waveforms. Finally, we note that separating the code sequences using differing chirp rates stabilised the overall performance of MDMA schemes but still required careful design of the waveforms for use in practical radar systems.

%=======================================================================
\section{Acknowledgements}
%=======================================================================

The computations described in this paper were performed using the University of Birmingham's BlueBEAR HPC service. See \href{http://www.birmingham.ac.uk/bear}{http://www.birmingham.ac.uk/bear} for more details. The authors would like to thank Emma-Claire Gurney for assistance with BlueBEAR.

\vfill\pagebreak
\bibliographystyle{IEEEbib}
\bibliography{ref}

\end{document}